\documentclass[fleqn,usenatbib]{mnras}

\usepackage[T1]{fontenc}
\usepackage{adjustbox}
\usepackage{tablefootnote}

\DeclareRobustCommand{\VAN}[3]{#2}
\let\VANthebibliography\thebibliography
\def\thebibliography{\DeclareRobustCommand{\VAN}[3]{##3}\VANthebibliography}

\newcommand{\GG}[1]{}
\usepackage{graphicx}	% Including figure files
\usepackage{amsmath}	% Advanced maths commands
\usepackage{adjustbox}
\usepackage{arydshln}
\usepackage{longtable}
\usepackage{lscape}
\usepackage{newtxtext,newtxmath}
\title[The survey of PNe in M~31 VI. PN LOSV in inner-halo substructures]{The survey of planetary nebulae in Andromeda (M 31) VI. Kinematics of M~31 inner-halo substructures and comparison with major-merger simulation predictions}

\author[S. Bhattacharya et al.]{
Souradeep Bhattacharya,$^{1}$\thanks{E-mail: souradeep@iucaa.in}
Magda Arnaboldi,$^{2}$
Francois Hammer,$^{3}$
Yanbin Yang,$^{3}$
Ortwin Gerhard,$^{4}$
\newauthor
Nelson Caldwell,$^{5}$
Kenneth C. Freeman,$^{6}$
\\
$^{1}$Inter University Centre for Astronomy and Astrophysics, Ganeshkhind, Post Bag 4, Pune 411007, India\\
$^{2}$European Southern Observatory, Karl-Schwarzschild-Str. 2, 85748 Garching, Germany \\ 
$^{3}$GEPI, Observatoire de Paris, PSL Research University, CNRS, Place Jules Janssen, F-92190 Meudon, France \\
$^{4}$Max-Planck-Institut für extraterrestrische Physik, Giessenbachstraße, 85748 Garching, Germany \\
$^{5}$Harvard-Smithsonian Center for Astrophysics, 60 Garden Street, Cambridge, MA 02138, USA \\
$^{6}$Research School of Astronomy and Astrophysics, Mount Stromlo Observatory, Cotter Road, ACT 2611 Weston Creek, Australia \\
}

\date{Accepted XXX. Received YYY; in original form ZZZ}

\pubyear{2021}

\begin{document}
\label{firstpage}
\pagerange{\pageref{firstpage}--\pageref{lastpage}}
\maketitle

\begin{abstract}
{M~31} has experienced a {recent} tumultuous merger history as evidenced from the many substructures that are {still} present in its inner halo, particularly the G1-Clump, NE- and W- shelves, and the Giant Stream (GS). We present planetary nebulae (PNe) line-of-sight velocity (LOSV) measurements covering the entire spatial extent of these four substructures. {We further use predictions for the satellite and host stellar particle phase space distributions for} a major merger (mass ratio = 1:4) simulation to help interpret the data. The {measured} PN LOSVs for the {two shelves} and GS are consistent with those from {red giant branch} stars. Their projected radius vs. LOSV phase space, links the formation of these substructures in a single unique event, consistent with a major merger. We find the G1-clump to be {dynamically cold compared to the M~31 disc} ($\rm\sigma_{LOS, PN}=27$~km~s$^{-1}$), consistent with {pre-merger} disc material. Such a structure can not form in a minor merger (mass ratio $\sim$1:20), and is therefore a smoking gun for the recent major merger event in M~31. {The simulation also predicts the formation of a predominantly in-situ halo from splashed-out pre-merger disc material, in qualitative agreement with observations of a metal-rich inner halo in M~31.}
Juxtaposed with previous results for {its discs}, we conclude that M~31 has had a recent (2.5--4~Gyr ago) `wet' major merger with the satellite falling along the GS, heating the pre-merger disc to form the M~31 thicker disc, {rebuilding the M~31 thin disc, and creating the aforementioned
inner-halo substructures. } 
\end{abstract}


%%%%%%%%%%%%%%%%%%%%%%%%%%%%%%%%%%%%%%%%%%%%%%%%%%

%%%%%%%%%%%%%%%%% BODY OF PAPER %%%%%%%%%%%%%%%%%%

\section{Introduction}
\label{sect:intro}
At a distance of $773$~kpc \citep{Conn16}, Andromeda (M~31; V$\rm_{sys}= - 301$~km~s$^{-1}$; \citealt{Watkins13}) is the nearest giant spiral galaxy to our Milky Way (MW), with its halo covering over 100~sq.~deg. on the sky. Photometric surveys \citep[e.g. PAndAS;][]{mcc09,mcc18} measuring the projected number density and photometric metallicity of resolved red giant branch (RGB) stars have revealed several faint and diffuse substructures in the M~31 halo. The most striking features in the inner halo include the Giant Stream (GS; a long trail of stars to the south-east of the M~31 disc spanning over $\sim$20~sq.~deg. on the sky; {see} \citealt{Ibata01}), the NE- and W-shelves (shell-like overdensities to the north-east and south-west of the M~31 disc; {see \citealt{Ferguson02} and \citealt{Fardal07} respectively}) and the G1-Clump (a distinct clump of stars at the southern tip of the M~31 disc; {see} \citealt{Ferguson02}). Such substructures are fossil records of hierarchical mass accretion in galaxies and their kinematic properties can shed light on the {mass, orbital parameters and other associated properties} of the accreted satellites \citep{Johnston08}.

However, owing to the proximity of M~31 and large {angular extent} of these substructures {on the sky}, spectroscopic observations of absorption lines in their RGB stars and subsequent line-of-sight velocity (LOSV) measurement have {until recently} been limited to small fields within the substructures, through pencil-beam spectroscopic surveys \citep[e.g.][]{Guha05}. Such surveys have been extensively carried out in the GS \citep{Ibata04,Guha05,Gilbert09}. The presence of two kinematic components (each having $\rm \sigma_{LOS} \sim 20$~km~s$^{-1}$ but offset by $\sim$100~km~s$^{-1}$) in small Keck/DEIMOS fields in the GS was first revealed by \citet{Gilbert09}. \citet{Escala22} measured the LOSVs of RGB stars in the NE-Shelf and found a 'wedge' pattern of the NE-Shelf RGB stars in their projected radial distance from galaxy center vs. LOSV phase space {(hereafter R$\rm_{proj}$ vs. V$\rm_{LOS}$ diagram)}. The recent spectroscopic survey of RGB stars in M~31 with the DESI multi-object spectroscopy instrument \citep{Dey22} also identifies the same pattern in the NE- and W-shelves as well as the two kinematic components in the GS but out to larger distance from the M~31 center. \citet{Reitzel04} obtained LOSV velocities of RGBs in the vicinity of the G1-clump but attributed their obtained measurements to the outer disc members of M~31, {rather than to the main G1-clump substructure, though an association of the clump with the nearby G1 globular cluster was ruled out}. The {general properties} of LOSV measurements {for the RGB stars} in the M~31 substructures {are} discussed in detail in \citet{fm16}. 

Dynamical simulations aiming to uncover the origin of these substructures, and thereby the recent merger history of M~31, have primarily been constrained by observed photometric properties (metallicity and surface brightness; see \citealt[][]{mcc18} for details) while comparing with specific predictions {only when} pencil beam spectroscopic measurements were available. \citet{Ibata04} first constrained the orbit of the GS, suggesting that the progenitor of the stream passed very close to the M~31 disc, causing its demise as a coherent structure and producing a fan of stars polluting the inner halo. N-body simulations of a `dry' merger along the GS placed the mass of the infalling satellite at $\sim 3.16 \times 10^{9}$~M$_{\odot}$ \citep[similar to the mass of the Large Magellanic Cloud;][]{Fardal06,Fardal13} {with the satellite also producing} the NE- and W- shelves as projected shells of disrupted satellite material following the merger \citep{Fardal07,Fardal12}. {The three dimensional shells which arise from the simulations of such an accretion event reproduce the LOSVs of planetary nebulae (PNe) in the M31 disc which are classified as outliers by \citet{Merrett03}.} \citet{Fardal13} further find that the satellite would have fallen in $\sim760$~Myr ago with a total mass $\leq10^{10}$~M$_{\odot}$, including dark matter (DM). Given the lack of star-formation burst in the {GS} or elsewhere in M~31 at this time \citep{bernard12,bernard15,wil17}, they posited that this would have been a gas-poor merger event. 

{ While there is an overall similarity in the R$\rm_{proj}$ vs. V$\rm_{LOS}$ diagram for the observed NE-Shelf RGB stars in \citet{Escala22} with that predicted by \citet{Fardal13}, these N-body simulations predict a single rather cold component  ($\rm \sigma_{LOS} \sim 20$~km~s$^{-1}$) for GS in this R$\rm_{proj}$ vs. V$\rm_{LOS}$ diagram. This simulated property of the GS is in tension with the presence of two kinematic components found instead in the GS by \citet{Gilbert09}. Another discrepancy is associated with the G1-clump, which is not predicted as a substructure in these simulations.} 

{\citet{ham18} have proposed a gas-rich major merger (mass ratio $\sim$1:4) 2--3~Gyr ago in M~31 where a satellite infalls along the GS, forms the two shelves in multiple pericenter passages, while also forming the G1-clump from perturbed disc material as well as forming a counter giant stream (CGS) substructure on the side of the M~31 disc opposite to the GS. Their simulation is consistent with the two kinematic components identified by \citet{Gilbert09} in the GS and they predict a broad LOSV distribution for the complete GS region due to the debris left by the satellite in multiple pericenter passages. In this scenario, the pre-merger M~31 disc would be dynamically heated to form a thick disc while the merger would be accompanied by star-formation with the bulk of the cold gas rebuilding the thin disc in M~31. Any minor mergers (satellite--to--M31 mass ratio $<$1:10) would not have any measurable impact on the M~31 disc kinematics \citep[E.g.][]{Hopkins09,Martig14}.}

{We thus turn to assess the observed effects of the recent merger on the M~31 disc. In \citet{Bh+19b}, we utilised observations of PNe in the M~31 disc to identify the kinematically distinct thin and thicker discs in M~31, traced by younger ($\sim2.5$ Gyr old) high-extinction PNe and older ($>4.5$ Gyr old) low-extinction PNe respectively. The thin and thicker discs, at the position of the solar neighbourhood in M~31, were found to respectively have velocity dispersions almost twice ($\sigma_{\phi}\sim61$~km~s$^{-1}$) and thrice ($\sigma_{\phi}\sim101$~km~s$^{-1}$) that of the MW population of the same age (also consistent with \citealt{dorman15} for the thick disc RGB stars; $\rm\sigma_{LOS}\sim90$~km/s). Such a dynamically hot thick disc can only be produced in a major merger event \citep{Hopkins09}. The resulting age-velocity dispersion relation from PNe in the M~31 disc suggests a merger mass ratio $\sim$1:5.}

{The kinematically distinct discs are also chemically distinct \citep{Bhattacharya22}, with the thicker disc having flat radial oxygen and argon abundance gradients, typical of discs that have experienced a recent major merger \citep[E.g.][]{Zinchenko15}, and the thin disc having steeper gradients consistent with that of the MW thin disc, taking the difference in disc scale lengths into account. \citet{Arnaboldi22} found that the [Ar/H] vs [O/Ar] plane for emission line nebulae is analogous to the [Fe/H] vs [$\alpha$/Fe] plane for stars, constraining the chemical enrichment and star formation history of the parent stellar population. They showed that the older M~31 thicker disc PNe show monotonic chemical enrichment with an extended star formation history up to super-solar values. By contrast, the younger M~31 PN progenitors are formed from accreted metal-poor gas mixed with the relatively metal-rich pre-merger M~31 inter-stellar medium (ISM).} 

{The major merger in M~31 would thus have been one with a gas-rich satellite, i.e. it would have been a `wet merger'. This is also consistent with the observations of $\sim$2~Gyr old burst of star-formation measured from deep \textit{Hubble Space Telescope} (HST) photometry in the M~31 disc \citep{wil17}, and also in small fields in several inner-halo substructures \citep{bernard15} as well as in the outer warp region in the disc \citep{bernard12}. A recent major merger scenario is also consistent with the observation of a metal-rich inner halo in M~31 \citep[][]{dsouza18,DSouza18b,Escala23} as well as {with} the two kinematically distinct globular cluster populations found in the M~31 outer halo \citep{Mackey19}. The merging satellite would have been massive enough to perturb the pre-merger disc to form the M~31 thicker disc while the thin disc was re-formed afterwards from gas brought in by the satellite mixed with that already present in the pre-merger M~31 disc. However, the formation of inner-halo substructures in the same wet-merger event has not yet been clearly established.}

{There is therefore the opportunity to set constraints on the merger history of M~31, by comparing predictions from major merger simulations with LOSV measurements at the different locations in the M~31 inner halo substructures with substantial spatial coverage. LOSV measurements over the entire spatial scale spanned by each of the M~31 inner halo substructures are thus required. This allows us to compare their observed position-velocity phase space properties with the predictions from simulations by \citet{ham18} so as to investigate their formation in the same major-merger event that formed the thin and thicker discs of M~31.}

\begin{figure*}
\centering
	\includegraphics[width=\textwidth,angle=0]{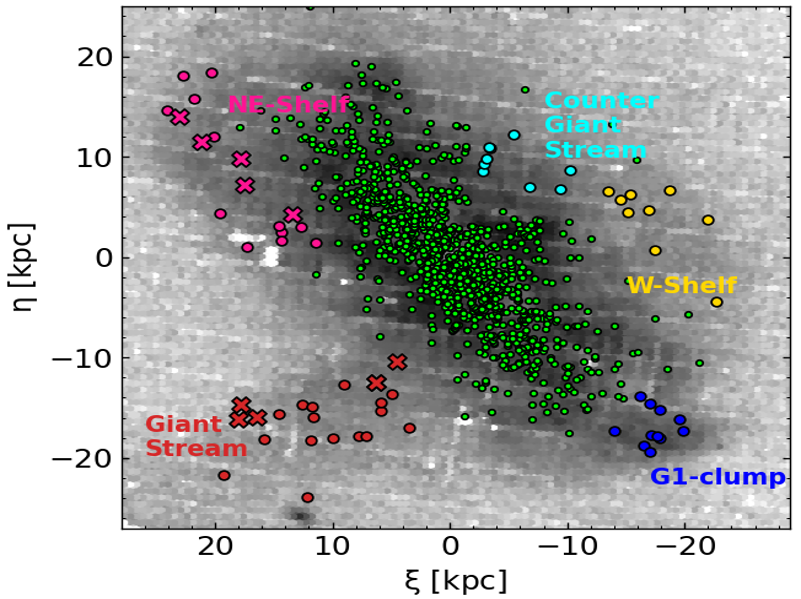}
	\caption{The number density map of RGB stars from PAndAS \citep{mcc18}, deprojected from their on-sky position and binned for visual clarity are shown in grey. {The PNe with LOSV measurements \citep{Bhattacharya22} are marked in green.} Those PNe co-spatial with the five marked M~31 inner-halo substructure regions studied in this work are marked with different colours. The Keck/Deimos field positions where RGB star LOSVs were measured for the NE-Shelf and GS are marked as crosses. The origin is at the center of M~31.}
	\label{fig:spat}
\end{figure*}

PNe, {the} emission-line nebulae in the late stages of stellar evolution, {are} established discrete tracers of {the} stellar population properties such as light, kinematics and chemistry in a number of galaxies of {different} morphological types \citep[e.g.][]{Aniyan18,Hartke22}. {PNe evolve from stars with initial masses between $\sim$0.8 and 8 M$\rm_\odot$, thus covering a wide range of parent stellar population ages ($>11$~Gyr to $\sim100$~Myr old). However, owing to the short visibility lifetime of PNe, only some stars in a stellar population are in the PN phase at any point in time. The number of identified PNe in any galaxy is linked to the total luminosity of the parent stellar population via a specific frequency that is related to the galaxy morphological type \citep{buz06}.} {Thus PN} LOSV measurements are simply a discrete sampling of stellar LOSV measurements in any galaxy, consistent with similar measurements from resolved stellar populations \citep[E.g.][]{Bh+19b} and integral-field spectroscopy \citep[E.g.][]{pul18,Aniyan21}. {The kinematic properties of the entire parent stellar population in a galaxy is effectively sampled by PNe even if multiple overlapping populations of different ages and kinematics are present \citep[E.g.][]{Aniyan18}. }

PN LOSV measurements are obtained from bright emission lines, which at the distances of M~31 can be carried out in reasonable allocation of telescope time with current instrumentation \citep{merrett06, Bhattacharya22}. Despite the large spatial extent of M~31 and its inner halo substructures, PN LOSV measurements can be carried out to effectively sample LOSVs from entire substructure regions. The PN kinematics smooth over field-to-field fluctuations that affect pencil-beam RGB LOSV measurements. {Furthermore, the MW halo is effectively transparent in the narrow-band [\ion{O}{III}] 5007~\AA~filter that is used for imaging PNe in M~31.} 

In this paper, we present in Section~\ref{sect:data}, the first LOSV measurements of PNe co-spatial with {the entire observed range of} M~31 inner-halo substructures in M~31, namely G1-Clump, W-Shelf, NE-Shelf and GS. {We also present the LOSV of the} PNe in the spatial region where the CGS is predicted by \citet{ham18}. {For the G1-clump, this is the first set of LOSV measurements over the entire spatial coverage of this substructure from any tracer}. We compare the observed LOSVs for each substructure with that of simulated analogues in the highest resolution major-merger (mass-ratio $\sim$1:4) simulation by \citet{ham18}. We compare the measured LOSV distribution and position-velocity plots in Section~\ref{sect:losvd}. We discuss our results, comparing also with other simulation predictions, in Section~\ref{sect:disc}. We conclude in Section~\ref{sect:conc}.

%____________________________________________________________

\section{Data and sample selection}
\label{sect:data}

\subsection{Observations}
\label{sect:obs}

The PNe were first identified in a 54 sq. deg imaging survey of M~31 with MegaCam at the CFHT (central 16 sq. deg.-- \citealt{Bh+19}; {outer fields} -- \citealt{Bh21}; {see also \citealt{PhDthesis}}). Spectroscopic observations of a complete subsample of these PN candidates were carried out with the Hectospec multifibre positioner and spectrograph on the Multiple Mirror Telescope \citep[MMT;][]{fab05}. Details of the spectroscopic observations and data reduction {are} discussed in \citet{Bhattacharya22}. Combined with the re-analysis of Hectospec spectra of archival PNe studied by \citet{san12}, we measure the LOSV (V$\rm_{LOS}$) of 1251 unique PNe (termed the \textit{PN\_M31\_LOSV} sample in \citealt{Bhattacharya22} {and plotted in Figure~\ref{fig:spat}}) from the strongest emission-lines with an uncertainty of 3 km s$^{-1}$. Only PN with confirmed detection of the [\ion{O}{iii}] 4959/5007 ~\AA~ emission lines {are considered in this work}.

\begin{table}
\caption{Measured properties of the M~31 PNe in the \textit{PN\_M31\_LOSV} sample. Column 1: Sl. No. of the PN in this work. Following IAU naming conventions, each PN should be designated as SPNA$<$Sl. No.$>$. E.g. PN 4040 should be termed SPNA4040; names assigned by \citet{san12} are reported where applicable; Column 2--3: spatial position of the PN; {Column 4: The [\ion{O}{III}] 5007 ~\AA~ magnitude measured in \citet{Bh+19} and \citet{Bh21}; Column 5: LOSV of the PN}; Column 6: Substructure the PN is co-spatial with. }
\centering
\adjustbox{max width=\columnwidth}{
\begin{tabular}{cccccc}
\hline
Sl. No. & RA [J2000] & DEC [J2000] & m$_{5007}$ & V$\rm_{LOS}$ & Subs\\
 & deg & deg & mag & km s$^{-1}$ & \\
\hline
4040 & 	8.7133009 & 	39.9081349 & 	20.85 & 	-489 & 	G1 \\
4055 & 	8.8975791 & 	40.0276394 & 	24.08	& -406 & 	G1 \\
4372 & 	8.7881072 & 	39.4735326 & 	21.61 & 	-461 & 	G1 \\
4378 & 	8.6723442 & 	39.5793943 & 	23.9 & 	-469 & 	G1 \\
4380 & 	9.0525232 & 	39.5948767 & 	21.27 & 	-445 & 	G1 \\
4381 & 	8.7473006 & 	39.5985174 & 	22.76 & 	-469 & 	G1 \\
4387 & 	8.4966837 & 	39.7048166 & 	22.25 & 	-489 & 	G1 \\
4409 & 	8.7355548 & 	39.4036082 & 	25.77 & 	-500 & 	G1 \\
4414 & 	8.807157 & 	39.957371 & 	25.37 & 	-505 & 	G1 \\
4643 & 	8.6981518 & 	39.6033913 & 	20.71 & 	-467 & 	G1 \\
4660 & 	8.5407341 & 	39.8254211 & 	23.92 & 	-460 & 	G1 \\
1145 & 	9.3681118 & 	42.1826747 & 	22.85 & 	-237 & 	WS \\
1162 & 	9.2896012 & 	42.2549368 & 	23.29 & 	-221 & 	WS \\
1241 & 	9.2808607 & 	42.0545865 & 	22.17 & 	-325 & 	WS \\
1243 & 	8.6005598 & 	42.0871777 & 	22.36 & 	-321 & 	WS \\
1247 & 	8.9614551 & 	42.3511456 & 	21.35 & 	-303 & 	WS \\
3542 & 	8.4065861 & 	41.2020724 & 	24.5 & 	-324 & 	WS \\
3555 & 	8.998651 & 	41.6796182 & 	22.46 & 	-201 & 	WS \\
pn1022 & 	9.113125 & 	42.1060828 & 	21.35 & 	-15 & 	WS \\
pn1031 & 	9.4765417 & 	42.2470283 & 	23.23 & 	-334 & 	WS \\
322 & 	13.1282865 & 	42.4818166 & 	24.22 & 	-322 & 	NES \\
373 & 	13.2746797 & 	42.7025819 & 	21.21 & 	-243 & 	NES \\
1268 & 	12.3680211 & 	40.9982886 & 	21.43 & 	-258 & 	NES \\
1291 & 	12.1182029 & 	41.2237641 & 	22.13 & 	-285 & 	NES \\
1301 & 	12.1552798 & 	41.288377 & 	23.06	 & -165 & 	NES \\
1306 & 	12.6611455 & 	41.3044376 & 	20.86 & 	-211 & 	NES \\
1667 & 	11.819722 & 	41.1784581 & 	25.15 & 	-311 & 	NES \\
1803 & 	11.9689563 & 	41.3185024 & 	25.69 & 	-118 & 	NES \\
4314 & 	12.8776729 & 	42.1161212 & 	21.26 & 	-266 & 	NES \\
4315 & 	13.3335287 & 	42.3056055 & 	22.27 & 	-288 & 	NES \\
4899 & 	13.048083 & 	42.7969448 & 	23.09 & 	-279 & 	NES \\
PNHK133 & 	12.10325 & 	41.1364175 & 	19.99 & 	-356 & 	NES \\
4169 & 	11.8471063 & 	38.8649003 & 	24.57 & 	-336 & 	GS \\
4173 & 	11.4665266 & 	38.9510126 & 	24.44 & 	-203 & 	GS \\
4177 & 	11.3043703 & 	39.0187884 & 	25.97 & 	-342 & 	GS \\
4178 & 	11.0967258 & 	39.0844815 & 	25.28 & 	-450 & 	GS \\
4180 & 	11.0388529 & 	39.1048582 & 	26.1 & 	-219 & 	GS \\
4181 & 	11.7784703 & 	39.1867887 & 	22.46 & 	-449 & 	GS \\
4182 & 	11.4938274 & 	39.2236968 & 	21.68 & 	-313 & 	GS \\
4184 & 	10.711797 & 	39.2753675 & 	26.07 & 	-388 & 	GS \\
4186 & 	11.6045208 & 	39.3359147 & 	26.08 & 	-133 & 	GS \\
4187 & 	11.5262993 & 	39.3408018 & 	22.83	& -380 & 	GS \\
4194 & 	10.9626204 & 	39.4210327 & 	24.26 & 	-264 & 	GS \\
4200 & 	10.9822732 & 	39.5134996 & 	25.85 & 	-366 & 	GS \\
4209 & 	10.9064413 & 	39.6348832 & 	22.97 & 	-488 & 	GS \\
4210 & 	11.3110249 & 	39.6547866 & 	25.81 & 	-178 & 	GS \\
4295 & 	12.0958552 & 	38.3689975 & 	24.53 & 	-400 & 	GS \\
4482 & 	11.3915279 & 	38.2884832 & 	25.07 & 	-245 & 	GS \\
1163 & 	10.568631 & 	42.2663786 & 	23.33 & 	-309 & 	CGS \\
1179 & 	10.5702728 & 	42.3418162 & 	23.64 & 	-225 & 	CGS \\
1190 & 	10.5618461 & 	42.3994257 & 	21.45 & 	-183 & 	CGS \\
1206 & 	10.5518751 & 	42.5260999 & 	22.01 & 	-303 & 	CGS \\
1208 & 	10.5638166 & 	42.5355199 & 	25.19 & 	-202 & 	CGS \\
1223 & 	10.3699999 & 	42.7097802 & 	22.74 & 	-281 & 	CGS \\
pn1028\tablefootnote{This PN was not observed by Megacam at the CFHT as it {lay} on a chip-gap. Hence no m$_{5007}$ value is reported here. It was observed spectroscopically by \citet{san12}.} & 	9.838166667 & 	42.42119611 & 	--	&  -332 & 	CGS \\
pn3008 & 	9.881625 & 	42.19908528 & 	23.2464 & 	-344 & 	CGS \\
pn3010 & 	10.14729167 & 	42.17555611 & 	22.16 & 	-143 & 	CGS \\
\hline
\end{tabular}
\label{table:data}
}
\end{table}

%____________________________________________________________

\begin{table*}
\caption{Properties of PNe in M~31 substructures. Column 1: Substructure name; Column 2: No. of PNe co-spatial with the substructure; Column 3--4: Mean V$\rm_{LOS}$ and $\rm\sigma_{LOS}$ of the PN in the substructure; Column 5--6: Mean V$\rm_{LOS}$ and $\rm\sigma_{LOS}$ of all simulated particles present in the substructure analogue; Column 7--8: Mean V$\rm_{LOS}$ and $\rm\sigma_{LOS}$ of the {host particles} in the simulated substructure analogue; Column 9--10: Mean V$\rm_{LOS}$ and $\rm\sigma_{LOS}$  of the {satellite particles} in the simulated substructure analogue; Column 11: The offset in projected radial distance (R$\rm_{proj, offset}$) of the simulated substructure from the observed substructure (positive values imply the simulated analogue is located further away); {Column 12: Fraction of particles in the simulated substructure analogue that originate from the host (f$\rm_{host}$)}.}
\centering
\adjustbox{max width=\textwidth}{
\begin{tabular}{cccccccccccc}
\hline
Substructure & N$\rm_{PN}$ & <V$\rm_{LOS, PN}$> & $\rm\sigma_{LOS, PN}$ & <V$\rm_{LOS, sim}$> & $\rm\sigma_{LOS, sim}$ & <V$\rm_{LOS, host}$> & $\rm\sigma_{LOS, host}$ & <V$\rm_{LOS, sat}$> & $\rm\sigma_{LOS, sat}$ & R$\rm_{proj, offset}$ & {f$\rm_{host}$}\\
 & & km s$^{-1}$ & km s$^{-1}$ & km s$^{-1}$ & km s$^{-1}$ & km s$^{-1}$ & km s$^{-1}$ & km s$^{-1}$ & km s$^{-1}$ & kpc\\
\hline
G1-Clump & 11 & $-469 \pm 8$ & $27$ & $-469$ & $62$ & $-476$ & $57$ & $-404$ & $69$ & 3.51 & 0.89\\
W-Shelf & 9 & $-253 \pm 34$ & $97$ & $ -354$ & $69$ & $ -357$ & $69$ & $ -349$ & $69$ & 1.06 & 0.63\\
NE-Shelf & 12 & $-258 \pm 19$ & $64$ &  $-271$ & $78$ & $-269$ & $70$ & $-277$ & $97$ & 0 & 0.73\\
{GS} & 16 & $-322 \pm 26$ & $102$ & {$-385$} & {$68$} & {$-348$} & {$68$} & {$-400$} & {$62$} & 7.47 & 0.28\\
CGS & 16 & $-258 \pm 24$ & $68$ & $-264$ & $88$ & $-269$ & $79$ & $-254$ & $105$ & 0 & 0.7\\

\hline
\end{tabular}
\label{table:prop}
}
\end{table*}

\subsection{Substructures in M~31 and their PN sub-samples}
\label{sect:subs_pn}

Figure~\ref{fig:spat} shows the map of RGB stars in the M~31 disc and inner halo from PAndAS \citep{mcc18}. The prominent substructures appear as overdensities in Figure~\ref{fig:spat} and are marked in different colours (GS -- red; NE-Shelf -- pink; G1-Clump -- dark blue; W-Shelf -- yellow). %In \citet{Bh21}, we investigated the properties of the PN Luminosity Function (PNLF) for the stellar populations in the disc and inner-halo substructures of M~31. From their PNLFs, we find that the GS and NE-Shelf are consistent with being composed of stellar debris from an in-falling satellite, while the G1-Clump is linked to the pre-merger M~31 disc with a significant contribution of younger stars presumably formed out of the accreted cold gas.
As in \citet{Bh21}, the PNe co-spatial within a given substructure are considered to belong to that substructure (also coloured accordingly) and used for subsequent analysis. We further mark the expected position of the counter-giant stream (CGS; cyan), a substructure predicted by the simulations of \citet{ham18} but not observed in the map of RGB stars from PAndAS (discussed further in Section~\ref{sect:sim}).  

In Table~\ref{table:data}, we present the V$\rm_{LOS}$ of the PNe in each of the substructures. We note that in \citet{Bh21}, we identified a larger number of fainter PNe in these substructures {from photometry} but the LOSV measurements were only possible for a {brighter} magnitude-limited sub-sample of the PNe. Additionally, Figure~\ref{fig:spat} shows the position of Keck/Deimos fields where RGB star LOSVs were measured for the NE-Shelf and GS by \citet{Escala22} and \citet{Gilbert09} respectively. The detailed comparison of the kinematics of the major merger simulations with the RGB observations of \citep{Dey22} will be addressed in a forthcoming publication.

%__________________________________________________________

\subsection{Observed line-of-sight velocity distribution in M~31 substructures}
\label{sect:losv_pn}

\begin{figure}
	\centering
	\includegraphics[width=\columnwidth,angle=0]{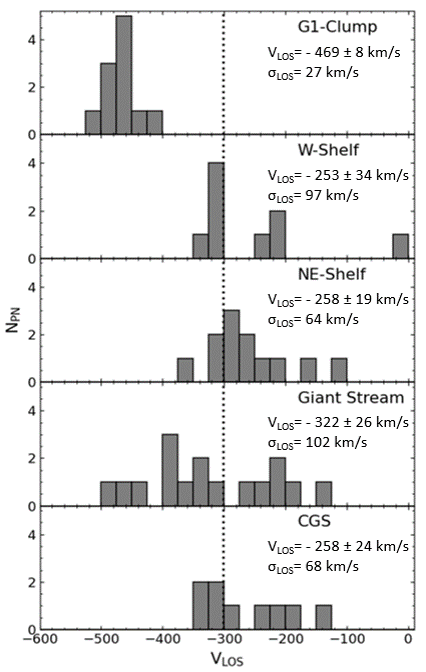}
	\caption{The LOSV distribution for five substructures in the M~31 inner halo from the observed PN measurements. {The dotted black line shows the systemic velocity of M~31. See Section~\ref{sect:losv_pn} for discussion.}}
	\label{fig:losv_pn}
\end{figure}

The number of PNe co-spatial with each substructure, their mean LOSV (V$\rm_{LOS, PN}$) and their LOSV dispersion ($\rm\sigma_{LOS, PN}$), are noted in Table~\ref{table:prop}. %We can see that these substructures are all relatively dynamically hot with even the coldest among them (G1-clump) having velocity dispersion similar to that of the MW thick disc \citep{Nord04}. 
Figure~\ref{fig:losv_pn} shows the LOSV distributions of PNe in the five substructure regions. We discuss each of them as follows:
\begin{itemize}
    \item G1-Clump: It has a narrow LOSV distribution, i.e, it is observed to be a {relatively} dynamically cold substructure having $\rm\sigma_{LOS, PN}=27$~km~s$^{-1}$ and V$\rm_{LOS, PN}=-469 \pm 8$~km~s$^{-1}$. {While this is the first dedicated LOSV measurement of the stellar population of the G1-Clump, we can retrospectively see that the RGB LOSV measurements from \citet{Reitzel04} in the vicinity of the G1-clump, then attributed to the M~31 outer disc by those authors, was plausibly a measurement of RGBs in the G1-clump (they do not provide the RGB positions and individual velocities to allow a direct comparison) with their V$\rm_{LOS, RGB}=-451$~km~s$^{-1}$ and $\rm\sigma_{LOS, RGB}=27$~km~s$^{-1}$, remarkably consistent with the values measured here.}
    \item W-Shelf: Given the faint nature of this substructure only 9 PNe with LOSV measurements are available. They distribute around {two} distinct LOSV values of $\sim -330$~km~s$^{-1}$ and $\sim -230$~km~s$^{-1}$ {with 5 and 3 PNe respectively (see Figure~\ref{fig:losvd}), and one additional PN at $\sim 0$~km~s$^{-1}$.} No RGBs spectra are available co-spatial with the W-shelf\footnote{Pencil-beam spectroscopy of RGB stars in the W-shelf were attempted by \citet{Fardal13} but their fields were along the M~31 minor axis, offset from the highest density of the W-shelf where we sample our PNe and close to the region at the boundary of the W-Shelf and CGS. Though some W-shelf stars would be present in these fields, they are more dominated by M~31 disc and halo stars with possible contribution from CGS stars. Hence we do not compare the PN LOSV properties with that of these RGBs.}. The R$\rm_{proj}$ vs. V$\rm_{LOS}$ diagram properties of the PNe and RGB from the DESI survey \citep{Dey22} in the W-shelf are discussed in Section~\ref{sect:proj}.
    \item NE-Shelf: The PNe in the NE-Shelf have $\rm\sigma_{LOS, PN}=64$~km~s$^{-1}$ and V$\rm_{LOS, PN}=-258 \pm 19$~km~s$^{-1}$. Their LOSV distribution is also broad though narrower than the GS. All NE-shelf RGBs studied by \citet{Escala22} have $\rm\sigma_{LOS, RGB}=89$~km~s$^{-1}$ and V$\rm_{LOS, RGB}=-284 \pm 4$~km~s$^{-1}$. The R$\rm_{proj}$ vs. V$\rm_{LOS}$ diagram properties of the PNe and RGB in the NE-shelf are discussed in detail in Section~\ref{sect:proj}.
    \item GS: The GS PNe have a broad LOSV distribution with $\rm\sigma_{LOS, PN}=102$~km~s$^{-1}$ and V$\rm_{LOS, PN}=-322 \pm 26$~km~s$^{-1}$. The two components in the GS, as identified by \citet{Gilbert09} {having $\rm \sigma_{LOS} \sim$20~km~s$^{-1}$ but offset by $\sim$100~km~s$^{-1}$ with V$\rm_{LOS, RGB}$ varying with distance from M~31 center (see additional discussion in Section~\ref{sect:proj})}, are not readily identifiable in the PN LOSV distribution (possibly given our lower number statistics). If taken together, all GS RGBs studied by \citet{Gilbert09} have $\rm\sigma_{LOS, RGB}=95$~km~s$^{-1}$ and V$\rm_{LOS, RGB}=-425 \pm 12$~km~s$^{-1}$. The lower mean LOSV of the RGB compared to the PN is because the RGB measurements in the GS are much closer to the M~31 disc than the PNe ({see} Section~\ref{sect:proj}). Note that the two kinematic components are also identified by DESI \citep{Dey22} at larger distances (also discussed further in Section~\ref{sect:proj}).

\begin{figure*}
	\centering
	\includegraphics[width=\textwidth,angle=0]{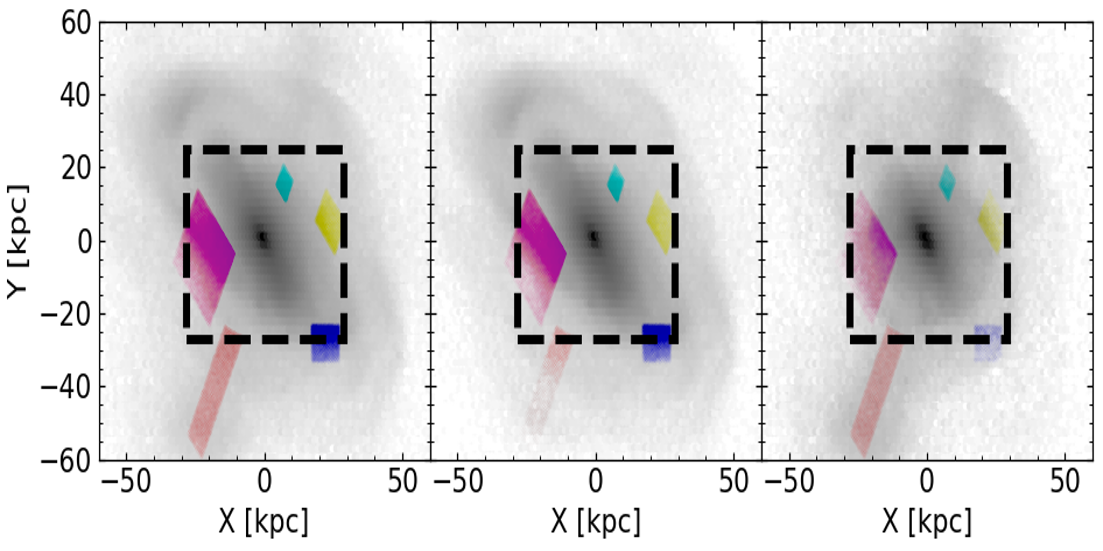}
	\caption{{For the major merger model in M~31, we show} the number density map of [left] all simulated particles, [middle] only host particles and [right] only satellite particles, accounting for the inclination of M~31. Particles in the five substructure analogous regions are marked with different colours, corresponding to that in Figure~\ref{fig:spat}, i.e., NE-Shelf -- pink, GS -- red, G1-clump -- blue, W-Shelf -- yellow, and CGS -- cyan . The dashed black box corresponds to the size of the M~31 field shown in Figure~\ref{fig:spat}.}
	\label{fig:spat_sim}
\end{figure*}
    
    \item CGS: This substructure is predicted to form by \citet{ham18} but not observed as an overdensity of RGB stars \citep{mcc18}. {The surface brightness of this substructure is predicted to vary with time (see Section~\ref{sect:sim} for details). Thus despite the non-observation of an RGB overdensity, the substructure may possibly be observed in the kinematic phase space.} The PNe here are selected around the {identified expected} position of the CGS and have $\rm\sigma_{LOS, PN}=68$~km~s$^{-1}$ and V$\rm_{LOS, PN}=-258 \pm 24$~km~s$^{-1}$. 
\end{itemize}

%____________________________________________________________

\section{Kinematics of simulated substructures in a major-merger scenario}
\label{sect:losvd}
In this section, we utilise the major-merger simulations from \citet{ham18} to help {with the interpretation of} the new LOSV data obtained for PNe and {for} the RGBs {which became} available in the literature for the substructures in the inner halo of M31. {We reiterate that PN LOSV measurements are a discrete sampling of their parent stellar population LOSVs, and thus are representative of the LOSV measurements of the entire M~31 stellar population. I.e., each PNe is representative of a the LOSV of a certain mass of stars in M~31 (to be precise, it is representative of a certain luminosity; see \citealt{buz06} for details). Since hydrodynamic simulations also have particles representative of a certain mass of stars, the PN LOSV distribution can directly be compared to that of any simulation for the same galaxy (as the number of particles would only be different from the number of PNe by a constant scale factor).}

%____________________________________________________________

\subsection{Major-merger simulation of M~31}
\label{sect:sim}

The merger simulations by \citet{ham18} were {tuned to} simultaneously {reproduce the} observed M~31 disc properties along with the spatial distribution of the faint inner-halo substructures. They were constrained by the structural properties of the inner M~31 galaxy, i.e, bulge-to-total ratio, the 10~kpc star-forming ring, bar and disc size, and the \ion{H}{i} disc, as well as by the {high velocity dispersion measured ($\rm\sigma_{LOS}\sim90$~km~s$\rm^{-1}$) for $\sim4$~Gyr old RGB stars in the M~31 disc} \citep[][]{dorman15}, and the burst of star-formation $\sim2$~Gyr ago observed throughout the M~31 disc and inner-halo substructures \citep{bernard15, wil17}. 

Assuming the GS is the trailing debris of the in-falling satellite (as is also the case for the minor merger scenario; \citealt{Fardal13}), and the NE- and W-shelves are formed after the merger from satellite debris, the orbital inclination of the satellite with respect to the M~31 pre-merger disc is constrained. Given the burst of star formation posits {a timing of the} event ~2-3 Gyr ago and given that a satellite massive enough to heat the pre-merger M~31 disc to the observed velocity dispersion values are necessary, \citet{ham18} model a bulgeless thin disc gas-rich (20\% gas fraction) satellite in-falling into a more massive gas-rich (20\% gas fraction) bulgeless thin disc which is the M~31 progenitor. The same DM fraction of 80\% is assumed for both the M~31 disc and satellite progenitors. 

The satellite infalls with its first pericentric passage $\sim$7~Gyr ago at a large distance from the M~31 disc. The second pericentric passage occurs $\sim$3--4~Gyr ago along the GS with the satellite losing some of its mass forming the NE- and W-shelves. However, the satellite is not {completely disrupted} and instead has a third pericentric passage that occurs $\sim$2--3~Gyr ago, forming an overdensity at the CGS and {depositing} further disrupted material in the GS. The GS formed in this process {contains} debris trails from the satellite but also of particles from the pre-existing disc. This {therefore} predicts a broad LOSV distribution for the GS. The satellite then has many short passages around the M~31 center that heats the pre-existing M~31 disc {stars to form a thicker disc having high velocity dispersion (as also measured in observations by \citealt{Bh+19b})} while also depositing satellite debris. The satellite core finally coalesces with the M~31 core to build-up the M~31 {central regions, leaving no separate remnant core}. The NE- and W-shelves also have more material deposited in multiple passages though the bulk of the material there is deposited in the second and third pericentric passages. The G1-clump is a dynamically cold substructure formed from perturbed material from the pre-merger M~31 disc and {is} thus predicted to {contain} metal-rich stars \citep{Faria07,Bh21}. \citet{ham18} find that a 1:4 mass ratio merger best reproduces the observed M~31 features they compared.

\begin{figure}
	\centering
	\includegraphics[width=\columnwidth,angle=0]{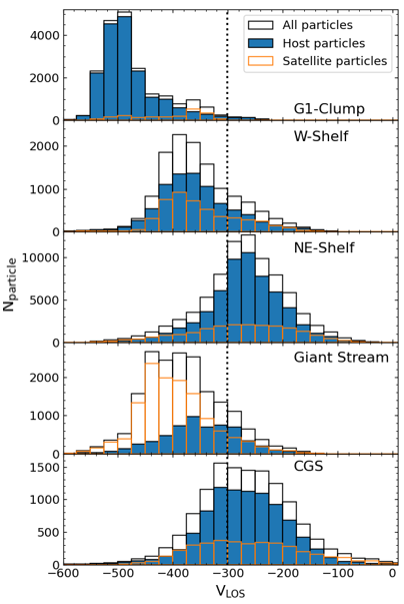}
	\caption{The LOSV distribution for the five substructures from their simulated analogues. The simulated particle LOSV distribution is shown for all (black unshaded), host (blue shaded) and satellite (orange unshaded) particles. {The dotted black line shows the M~31 $\rm V_{sys}$ value.}}
	\label{fig:losvd}
\end{figure}

{In this work, we utilise Model 336 from \citet{ham18} for comparing with observations as this is their best model\footnote{{Five more models have been marked as best models by \citet{ham18} but ran at lower resolution with only 2 million particles.}} for M~31 that was chosen for their highest resolution run with 20 million particles.} Figure~\ref{fig:spat_sim} shows the simulated spatial distribution of the as M~31 would appear from our line-of-sight. The three panels of Figure~\ref{fig:spat_sim} show all particles {(left)}, only host particles {(center)} and only satellite particles {(right)} respectively. The five {analogous regions cospatial} to the observed substructures are marked. {Both host and satellite particles are observed out to large radii (Figure~\ref{fig:spat_sim}). Their distribution is discussed further in Section~\ref{sect:halo}.} We note that the disc scale length of the M~31 analogue in this model is {3.6}~kpc, which is lower than that for the observed M~31 disc scale length of $\sim$6~kpc \citep{Yin09}. We note also that the substructures evolve over time (see also \citealt{cooper11}) and the model snapshot in Figure~\ref{fig:spat_sim} is chosen such that the surface brightness of the substructures are close to their observed values. Hence the positions of the model substructures are not exactly at the same distance from the M~31 center as their observed counterparts. The offset in projected radius of the simulated substructure from the observed ones, R$\rm_{proj, offset}$, has been noted in Table~\ref{table:prop}. We can see that the simulated NE-shelf and CGS subregions lie at the same projected radius as their observed analogues but the other substructures are {slightly} offset. The simulated G1-clump, W-shelf and GS occur $3.51$~kpc, $1.06$~kpc and $7.47$~kpc further away respectively than their observed analogues. 

The mean LOSV (V$\rm_{LOS, sim}$) and their LOSV dispersion ($\rm\sigma_{LOS, sim}$) for the particles in each simulated substructure are noted in Table~\ref{table:prop}. {We separately select those particles from the simulation which originate from the M~31 host (or were formed in the resulting disc after the merger) and those originating the satellite.} The mean LOSV and dispersion of host (V$\rm_{LOS, host}$, $\rm\sigma_{LOS, host}$) and satellite (V$\rm_{LOS, sat}$, $\rm\sigma_{LOS, sat}$) particles for each substructure are also listed in Table~\ref{table:prop}. 
%____________________________________________________________

\subsection{Line-of-sight velocity distribution of simulated substructures}
\label{sect:los}

{Figure~\ref{fig:losvd} shows the LOSV distributions of simulated particles (separately for all, only host and only satellite particles) for the substructure analogues. This can be qualitatively compared to the LOSV distributions of observed PNe\footnote{{We note that even though RGB star positions and velocities are available for the Keck/Deimos fields in the NE-Shelf and GS \citep{Gilbert09,Escala22}, the selection function for these RGB stars as well as the choice of field positions makes a direct statistical comparison of LOSV distributions complicated. An additional challenge is the contamination from the MW disc and halo RGBs that vary across field positions in M~31. This is further compounded by the offset in the positions between the observed substructures and their simulated analogues. They are useful for comparing with the simulations upon considering both the spatial and LOSV distributions as in \citet{Escala22}, discussed in Section~\ref{sect:proj}.}} in the M~31 inner halo substructures (Figure~\ref{fig:losv_pn}). }

{The G1-Clump  appears as a relatively dynamically cold substructure in the simulation, predominantly consisting of host particles that peak at a mean value consistent with observation (see Table~\ref{table:prop}) with satellite particles showing a distinct distribution. For the NE- and W- shelves, both host and satellite particles have overlapping LOSV distributions, with nearly the same mean LOSV value (see Table~\ref{table:prop}). The particle LOSV distributions in the NE-shelf span the same range as that of the observed PNe while for the W-shelf, a wider distribution is seen than the observed multi-modal LOSV distribution. The satellite and host particles of the GS also have overlapping distributions but have distinct mean LOSV values (see Table~\ref{table:prop}). The LOSV range spanned is similar to that of the observed PNe in the GS. For the CGS, much like the shelves, the LOSV distributions of host and satellite particles overlap with similar mean values though the satellite particles are expected to be dynamically hotter (see Table~\ref{table:prop}). The LOSV range spanned is similar to the observed PNe at the expected position of the CGS.}

%From Table~\ref{table:prop}, the observed mean LOSVs of the substructures from their PNe are consistent with the mean values of their simulated analogues within error  (except for the W-Shelf and GS).  It is also clear from Table~\ref{table:prop} that the mean LOSVs and dispersions of the host and satellite components of {some of the simulated substructures are different}. 

{For a quantitative analysis} to understand which component of the simulated substructure, if any, are consistent with the observed substructure kinematics, we statistically compare the observed PN LOSV distributions with that of their simulated analogues with an Anderson-Darling test (AD-test; \citealt{ADtest}), distinguishing between the host and satellite components. {The AD-test usage has been discussed in detail in \citet[][see Section~3.4]{Bh21}. It is adopted here to allow the comparison of numerically different data-sets.} In Table~\ref{table:comp}, we note the test significance (equivalent to a p-value) for each compared pair (observed vs. simulated particles separately for all, only host and only satellite particles) for each substructure. {The substructure LOSV distributions are identified as being drawn from different parent distributions if the significance is below 0.05. Otherwise, they may or may not be drawn from the same parent distribution.} We discuss the comparison for each of the substructures as follows:
\begin{itemize}
    \item G1-Clump: The observed LOSV distribution of the PNe and satellite particles are drawn from different parent distributions (see Table~\ref{table:comp}), indicating that the G1-clump is not formed from satellite debris. However, its observed LOSV distribution is consistent with having formed {mostly} from the host debris especially considering the simulated host mean LOSV, V$\rm_{LOS, host}=-476$~km~s$^{-1}$ is remarkably similar to the observed value. However, the simulated host LOSV dispersion is higher than observed with $\rm\sigma_{LOS, host}=57$~km~s$^{-1}$. 
    \item W-Shelf: The observed multi-modal LOSV distribution of the W-Shelf (Figure~\ref{fig:losv_pn}) is not observed for the simulated W-Shelf and hence the AD-test identifies that the simulated particles (also separately host and satellite ones) are drawn from a different parent distribution than the observed values.
    \item NE-Shelf: The LOSV distributions of both the satellite and host particles are consistent with the observed LOSV distribution.
    \item GS: The observed LOSV distribution of the PNe and host particles are drawn from different parent distributions (see Table~\ref{table:comp}) but the satellite particles are consistent with the observed LOSV distribution. %The host and satellite particles have similar mean LOSV {dispersion} but the satellite particles have {larger} LOSV dispersion (see Table~\ref{table:prop}).
    \item CGS: %There is an overlap of the LOSV distribution of the host and satellite particles in this simulated substructure, though the satellite particles are expected to be dynamically hotter (see Table~\ref{table:prop}). From the AD-test, 
    {The observed PN LOSV distribution is compatible with that for both host and satellite particles.} Hence the {presence of a distinct substructure at the location of the CGS cannot be established from the} LOSV distribution.
\end{itemize}

\begin{table}
\caption{AD-test significance comparing observed PN LOSV distribution of each substructure with that of their simulated analogues, separately for all particles, {host particles and satellite} particles. Note 0.25 and 0.001 are {set} as the maximum and minimum values respectively for the test significance. If the observed LOSV distribution has been drawn from a different parent distribution than the compared simulated LOSV distribution, they are italicized and marked with $^*$.}
\centering
\adjustbox{max width=\textwidth}{
\begin{tabular}{c|ccc}
\hline
Substructure & \multicolumn{3}{c}{Ad-test significance compared to simulation}\\
 & all particles & host particles & satellite particles \\
\hline
G1-Clump & 0.13 & 0.083 & \textit{0.001*}\\
W-Shelf  & \textit{0.001*} & \textit{0.001*} & \textit{0.001*}\\
NE-Shelf & 0.25 & 0.25 & 0.25\\
GS & 0.065 & \textit{0.049*} & 0.06\\
CGS  & 0.25 & 0.25 & 0.25\\
\hline
\end{tabular}
\label{table:comp}
}
\end{table}

While comparing the LOSV distributions of the observed and simulated substructures does shed light on their possible origin as disrupted material from either the satellite or the host galaxy, further insight may be obtained considering both the spatial and LOSV distributions. 

%____________________________________________________________

\subsection{Distribution of observed and simulated substructures in position-velocity phase space}
\label{sect:proj}

We compare the distribution of the observed PNe and simulated particles from \citet{ham18} in the R$\rm_{proj}$ vs. LOSV diagram. {Recall from Section~\ref{sect:sim}} that the simulated substructures do not appear at the {exact} same distance as their observed counterparts. So we subtract R$\rm_{proj, offset}$ {(see Table~\ref{table:prop})} from the projected distance of each simulated particle in a given substructure. We discuss the comparison for each of the substructures as follows:

\begin{figure}
	\centering
	\includegraphics[width=\columnwidth,angle=0]{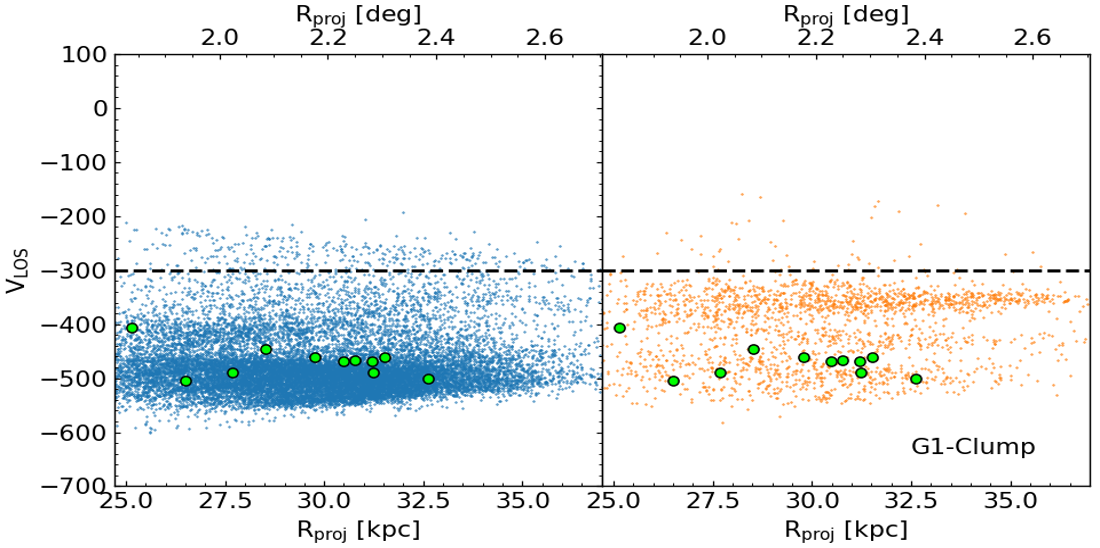}
    \caption{{The G1-Clump in the R$\rm_{proj}$ vs. V$_{LOS}$ diagram. The observed PNe are shown in green. The host and satellite particles are marked in blue and orange in the left and right panels respectively. On the lower x-axis distances are in units of kpc from center of M~31, while on the upper x-axis they are in units of degrees.  The projected distances of the simulated particles has $\Delta R\rm_{proj, offset} = 3.51$~kpc added to account for the slightly different positions of the simulated analogues from the observed substructure on the sky. The horizontal black dashed line shows the M~31 $\rm V_{sys}$.}}
	\label{fig:proj_G1}
\end{figure}

\begin{itemize}
    \item G1-Clump: {The R$\rm_{proj}$ vs. V$_{LOS}$ diagram for the simulated particles and PNe in the G1-clump are shown in Figure~\ref{fig:proj_G1}.} The bulk of the host particles in the simulated G1-Clump appear as an extended structure in R$\rm_{proj}$ with LOSV $\sim - 480$~km~s$^{-1}$. This structure is indeed dynamically cold. The observed PNe in the G1-Clump overlap with this host simulated analogue structure, albeit with slight offset in velocity. {While there are a few satellite particles also at the position of the PNe in Figure~\ref{fig:proj_G1}, the satellite particle LOSV distribution is clearly more uniform with an mean LOSV value of $-404$~km~s$^{-1}$. Given that PNe trace the kinematics of their parent stellar population and given that $\sim90\%$ of the stellar population in the G1-clump comes from the host material (Table~\ref{table:prop}), the PN thus trace the host material kinematics in M~31.} 

\begin{figure}
	\centering
	\includegraphics[width=\columnwidth,angle=0]{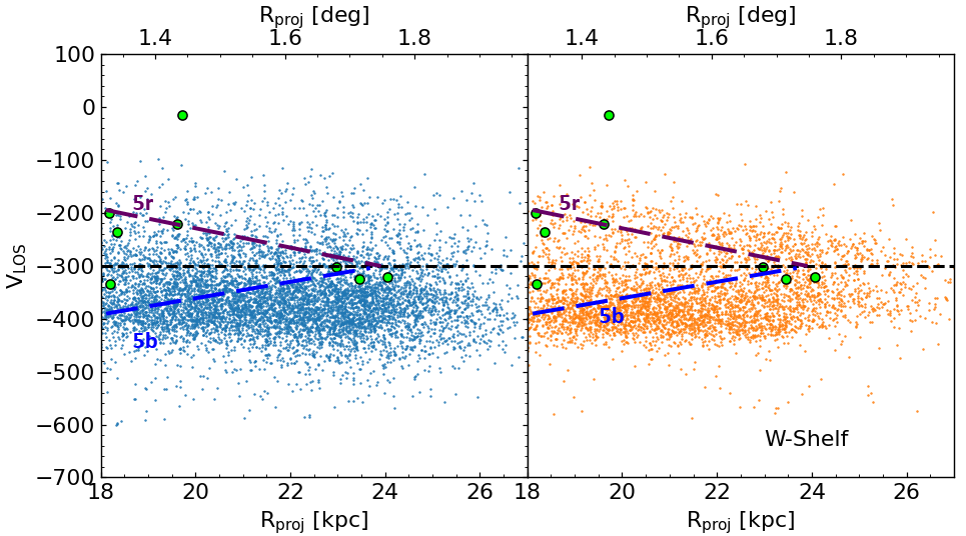}

	\caption{{Same as Figure~\ref{fig:proj_G1} but for the W-shelf. Additionally, the overdensity boundaries identified from the RGB stars in the DESI survey in this phase space are marked with dashed lines (see Section~\ref{sect:proj} for details).}}
	\label{fig:proj_WS}
\end{figure}

    \item W-Shelf: {The R$\rm_{proj}$ vs. V$_{LOS}$ diagram for the simulated particles and PNe in the W-Shelf are shown in Figure~\ref{fig:proj_WS}. An asymmetric wedge pattern with  the lower arm having higher particle density is visible for the satellite particles. The host particles occupy nearly the same parameter space as the lower arm but are denser at higher projected radii. The PNe overlap quite well with the upper arm of the satellite particle distribution, though there is only one PN present at an offset with the lower arm. The RGB stars from the DESI survey \citep{Dey22} also trace a wedge pattern in the W-shelf, with the overdensity boundaries (5r -- purple; 5b -- blue) marked in Figure~\ref{fig:proj_WS}. We note that the PNe, the RGBs\footnote{\citet{Dey22} also show that the PNe presented here (LOSV measured by those authors from archival spectra of our observations) trace the same feature as their RGB stars in position-velocity phase space.} and the W-shelf simulated satellite debris trace the same upper arm of the wedge. The RGB overdensity boundary position of the lower arm of the wedge, consistent with that of the single PN at that position, is also offset from the lower arm of the simulated satellite particles. }   

\begin{figure}
	\centering
	\includegraphics[width=\columnwidth,angle=0]{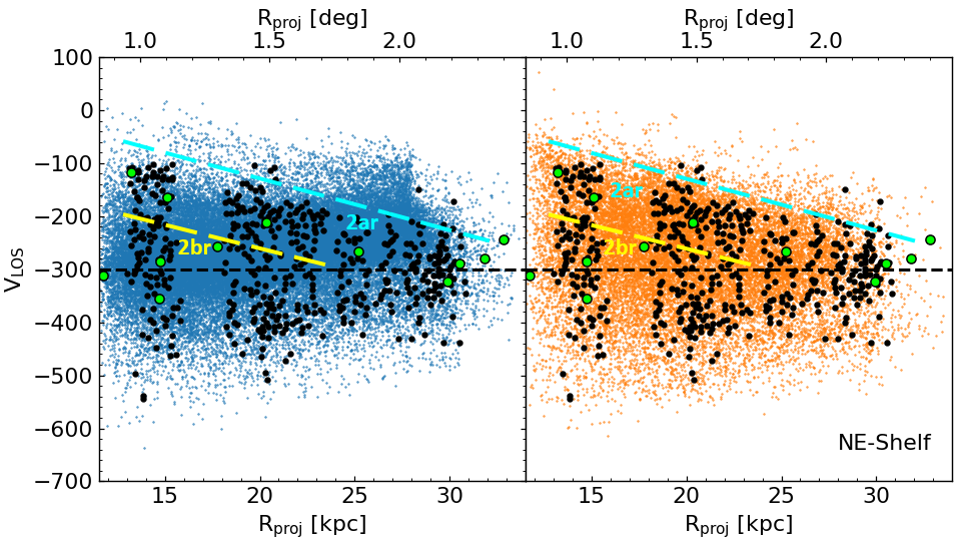}

	\caption{{Same as Figure~\ref{fig:proj_G1} but for the NE-shelf. The overdensity boundaries identified from the RGB stars in the DESI survey in this phase space are marked with dashed lines (see Section~\ref{sect:proj} for details). Additionally, the RGB stars identified by \citet{Escala22} are marked in black.} }
	\label{fig:proj_NES}
\end{figure}

\begin{figure}
	\centering
	\includegraphics[width=\columnwidth,angle=0]{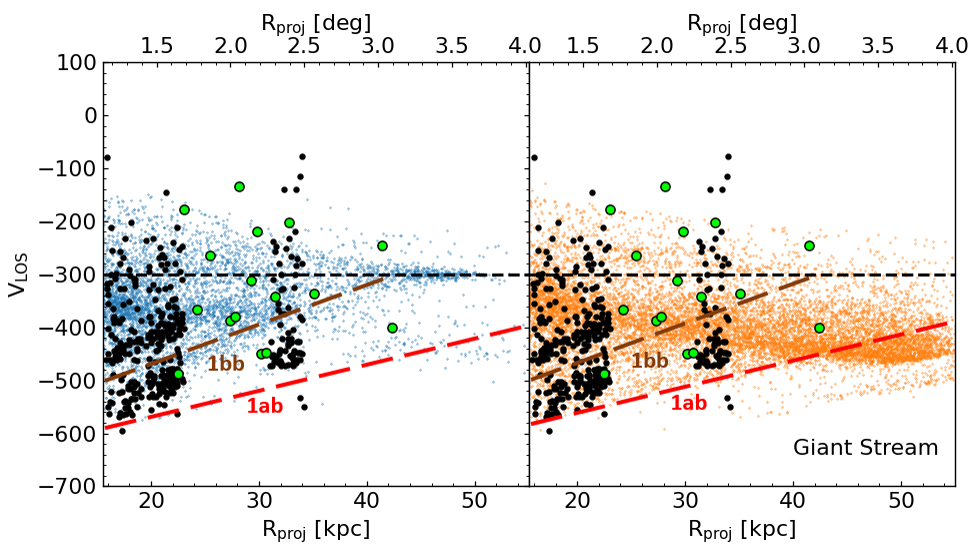}
 
	\caption{{Same as Figure~\ref{fig:proj_G1} but for the GS. The overdensity boundaries identified from the RGB stars in the DESI survey in this phase space are marked with dashed lines (see Section~\ref{sect:proj} for details). Additionally, the RGB stars identified by \citet{Gilbert19} are marked in black.}}
	\label{fig:proj_GS}
\end{figure}

    \item NE-Shelf: {The R$\rm_{proj}$ vs. V$_{LOS}$ diagram for the simulated particles and PNe in the NE-Shelf are shown in Figure~\ref{fig:proj_NES}. The simulated satellite particles show a filled wedge pattern, distributing not only along the arms of the wedge but rather occupying} a broader V$\rm_{LOS}$ range that reduces with increased R$\rm_{proj}$. The rotating M~31 thick disc is also visible from the distribution of the host particles {having V$\rm_{LOS}\sim-100$~km~s$^{-1}$ at R$\rm_{proj}>25$~kpc}. {The PNe are consistent with both the satellite and host particles in this figure.} The RGB stars from \citet{Escala22}, nearly co-spatial with the detected PN positions, overlap with the wedge pattern observed for the NE-shelf,{ though some of the RGB stars may be associated with the host particles that overlaps with the satellite material in this phase space.} Figure~\ref{fig:proj_NES} also {shows} the overdensity boundaries (2ar -- cyan; 2br -- {yellow}) in the R$\rm_{proj}$ vs. V$_{LOS}$ phase space of the NE-shelf identified from the RGBs in the DESI survey \citep{Dey22}. 2ar traces the upper boundary of the wedge traced by the RGBs from \citet{Escala22} as well as that from the satellite debris in the simulation. 2br is not clearly traced by the DESI RGBs (see Figure 6 in \citealt[][]{Dey22}) but its expected position is consistent with the PNe. {Overall there is no clear apparent} overdensity boundary in this phase space region for the RGBs observed by \citet{Escala22} nor in the satellite debris in the simulation. {We note that the lower boundary of the NE-shelf in this phase space is traced by the PNe cospatial with the M~31 disc but which are velocity outliers \citep{Merrett03}.}
    \item GS: {The R$\rm_{proj}$ vs. V$_{LOS}$ diagram for the simulated particles and PNe in the GS are shown in Figure~\ref{fig:proj_GS}.  The PNe co-spatial with the GS trace a relatively broad distribution in LOSV around the satellite particles. Their spread in LOSV is similar to that of the RGB stars from \citet{Gilbert09}.} Figure~\ref{fig:proj_GS} also marks the median position of the RGB overdensity boundaries in this phase space \citep{Dey22} identified recently in the DESI survey of M~31 (1ab -- red; 1bb -- purple). 1ab and 1bb are the same two kinematic components identified by \citet{Gilbert09} (see the overdensities for the RGBs nearest to these lines in Figure~\ref{fig:proj_GS} but with a small velocity offset) but traced out to larger radii. {There are some PNe that are along these tracks for the simulated host and satellite particles.} We discuss these features in comparison with the simulation in Section~\ref{sect:gs}. {We note that the main broad high-density nearly-horizontal track of satellite particles seen in Figure~\ref{fig:proj_GS} does have corresponding RGB stars observed at this position in the phase space (see Figure 6 in \citealt{Dey22}) but this is not marked as an overdensity by these authors.}  %We note here that the satellite particles in the GS trace a broader region in Figure~\ref{fig:proj} than the corresponding plot for the minor merger scenario seen by \citep[][see their Figure 15]{Kirihara17}. 

\begin{figure}
	\centering
 \includegraphics[width=\columnwidth,angle=0]{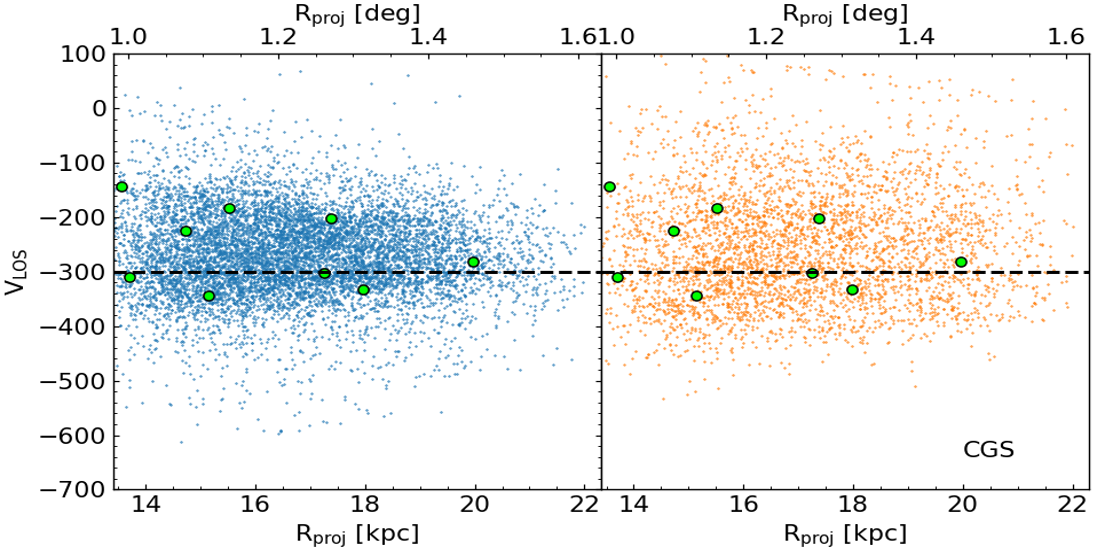}
	\caption{{Same as Figure~\ref{fig:proj_G1} but for the CGS.}}
	\label{fig:proj_CGS}
\end{figure}
    
    \item CGS: {The R$\rm_{proj}$ vs. V$_{LOS}$ diagram for the simulated particles and PNe in the CGS are shown in Figure~\ref{fig:proj_CGS}. While there is an overdensity of host particles (f$\rm_{host}\sim70\%$; Table~\ref{table:prop}) %that appears to be traced by the PNe
    , the satellite and host particles mostly overlap in Figure~\ref{fig:proj_CGS}.} As such the PNe can not clearly be attributed to either of the host or satellite. The predicted satellite particle distribution of the CGS is too sparse and close to the simulated M~31 disc. An exploration of PN LOSVs at larger radii along the predicted position of the CGS may be useful in identifying or denying its existence in the R$\rm_{proj}$ vs. LOSV phase space.
\end{itemize}

In summary, the R$\rm_{proj}$ vs. V$_{LOS}$ phase space, together with the LOSV distribution comparison in Section~\ref{sect:los} provides constraints on the origin of some of the inner-halo substructures in M~31. The {relatively dynamically cold ($\rm\sigma_{LOS, PN}=27$~km~s$^{-1}$)} G1-Clump substructure is consistent with {a perturbation of} the pre-merger M~31 disc, {which is a prediction of} the major merger simulation. {The RGB (consistent also with the PNe) trace the wedge structure for the W-Shelf in the R$\rm_{proj}$ vs. V$_{LOS}$ phase space, but are different from the asymmetric wedge shape traced by the satellite particles in the W-shelf simulated analogue.} The GS PNe have a broad distribution consistent with the R$\rm_{proj}$ vs. V$_{LOS}$ phase space positions predicted by the major merger simulation. The PNe co-spatial with the NE-Shelf and the CGS location may originate from either the host or the satellite. 

%____________________________________________________________

\section{Discussion}
\label{sect:disc}

\subsection{The G1-clump LOSV measurements as a smoking gun for the major merger in M~31}
\label{sect:g1}

Given the low luminosity of the G1-clump (L$_{tot, V-band}=9.37\times10^6$ M$_{\odot}$), \citet{Ferguson02} had initially suggested it to be a dissolving dwarf galaxy. However, its measured photometric metallicity ([M/H]$=-0.37$; \citealt{bernard15}) is too high for a dwarf galaxy. Thus, it has been suggested to be a remnant perturbed from the M~31 disc \citep{Faria07}. Minor mergers with mass ratio of the order 1:20 (as simulated by \citealt{Fardal13,Kirihara17}) do not perturb the disc of the host galaxy significantly enough to greatly increase its rotational velocity dispersion \citep{Martig14}, at least not even close to $\sim$3 times that of the MW as measured for the M~31 thicker disc \citep{dorman15, Bh+19b}. So producing a substructure from {perturbed} disc material, such as the G1-clump necessitates a major merger event. The G1-Clump has observed metallicity distribution, star formation history \citep{bernard15} and {PN Luminosity Function (PNLF)} consistent with that of the M~31 disc \citep{Bh21}. Its LOSV distribution and R$\rm_{proj}$ vs. V$_{LOS}$ diagram {are} consistent with {those} predicted by the major merger simulation by \citet{ham18}, that links its origin to the same merger event that formed the GS, NE-shelf and W-shelf. {No separate merger event for the formation of the G1-clump is needed.}

The G1-clump LOSV measurements are thus a smoking gun signature that show that the same wet merger \citep{Arnaboldi22} event that produced the kinematically \citep{Bh+19b} and chemically \citep{Bhattacharya22} distinct thin and thick disc in M~31, had the satellite in-fall along the GS (see orbital properties in Table 1 and associated discussion in \citealt{ham18}) producing the NE- and W-shelves from satellite debris material {while} the G1-clump {formed out} of perturbed disc material.

\subsection{The wedge pattern for the NE- and W-shelves}
\label{sect:shelves}

The wedge patterns in the R$\rm_{proj}$ vs. V$_{LOS}$ diagram {for the satellite particles} in the NE- and W-shelves are predicted by both minor \citep{Fardal13,Kirihara17} and major merger scenarios (Section~\ref{sect:proj}) in M~31 though the relative intensity of the arms of the {wedges} depend on the morphology and orbital inclination of the infalling satellite \citep[][]{Kirihara17}. %\citet{Fardal13} previously utilised spectroscopy of RGB stars along the minor-axis of M~31 and overlapping with the W-shelf overdensity to search for this pattern. While their measured LOSV values were consistent with their simulated values for the W-Shelf from a minor-merger model, they did not clearly observe the wedge pattern. 
The large spatial coverage of our PN survey allows us to map PNe belonging to different parts of the W-shelf and thereby, {in conjunction with the RGB surveyed in the DESI survey \citep{Dey22},} allowed us to observe its wedge pattern in the R$\rm_{proj}$ vs. LOSV phase space. %We thus find evidence of the 3D structure of the W-shelf which leads to the consequent wedge pattern. 
{The density of the W-shelf wedge pattern is asymmetric for the satellite particles in the major merger model (Figure~\ref{fig:proj_WS}) while the observations support a more symmetric wedge.}   

We do not observe the wedge-pattern clearly in the NE-Shelf solely from our PN kinematics. However, in tandem with the RGB LOSVs in the NE-Shelf from \citet{Escala22} (also \citealt{Dey22}), the wedge pattern in the R$\rm_{proj}$ vs. V$_{LOS}$ diagram is observed. The observed pattern is consistent with {the satellite debris for} both minor merger (see Figure~10 in \citealt{Escala22}) and major merger simulation predictions (Figure~~\ref{fig:proj_NES}). %The predicted NE-shelf satellite debris wedge pattern does depend on satellite-mass, morphology and orbital characteristics with the merger of a low-mass gas-poor spheroidal satellite making a more `pointed' wedge structure covering a small range of LOSV at higher R$\rm_{proj}$ (see Figure 10 in \citealt{Escala22}; $\rm\Delta$V$\rm_{LOS}\sim150$~km~s$^{-1}$ at R$\rm_{proj}\sim$30~kpc) compared to the `broader wedge end' produced in the major merger simulation (Figure~~\ref{fig:proj_NES}; $\rm\Delta$V$\rm_{LOS}\sim250$~km~s$^{-1}$ at R$\rm_{proj}\sim$30~kpc). \citet{Kirihara17} showed that disc and spheroidal satellite morphologies also produced subtly different satellite debris distributions in the R$\rm_{proj}$ vs. LOSV phase space for both the shelves. %Presence of a merger remnant in the NE-shelf, which would have resulted in an overdensity in the satellite debris and a `z' pattern in the R$\rm_{proj}$ vs. LOSV phase space \citep{Fardal08}, is ruled out as was also {found} by \citet{Escala22}. 
While some of the RGB stars and PNe are tracing the wedge pattern of satellite debris, some fraction of them are likely also tracing the M~31 thick disc that overlaps the NE-shelf in the R$\rm_{proj}$ vs. LOSV phase space (Figure~\ref{fig:proj_NES}). 

%It is noteworthy that the minor merger simulations from \citet{Fardal13} only include a thin disc for M~31, which is not consistent with observations \citep{dorman15,Bh+19b}. {It thus} can not be said that the RGB stars only trace the satellite particles in the wedge-pattern even in case of a minor merger, contrary to what was concluded by \citet{Escala22}. 

{In this regard, we note that f$\rm_{host}\sim73\%$ and $\sim63\%$ for the NE-shelf and W-shelf simulated analogues respectively. Since PNe trace the kinematics of the entire parent stellar population, it is expected that similar percentages of PNe would trace the NE- and W-shelf host LOSVs. But at least for the W-shelf, the PNe seem to only trace the satellite particles (Figure~\ref{fig:proj_WS}). A feature of the major merger simulations in \citet{ham18} is that the positions of the substructures vary with time. The large percentage of host particles in the shelves is likely a consequence of these substructures forming close to the M~31 disc in these simulations (see Section~\ref{sect:sim} for details), with the disc being thicker and dynamically hotter than that adopted in minor merger simulations \citep[e.g.][]{Fardal13}.}
%Physically these substructures must be further away from the M~31 disc. Nevertheless, the thick disc in these simulations is closer to observations than only the thin disc that is included in the M~31 minor merger simulations by \citet{Fardal13}. }

%Given the overlap of host and satellite particles in the R$\rm_{proj}$ vs. LOSV phase space (Figure~\ref{fig:proj_NES}), the PNe are consistent with both structures. However, for the W-shelf, the PNe and RGB overdensity lie on a wedge structure in this phase space (Figure~\ref{fig:proj_WS}), likely all tracing the satellite debris. They do not overlap with the highest density of the host particles in this phase space (Figure~\ref{fig:proj_WS}), corresponding with the extension of the M~31 disc in \citet{ham18}. It thus seems unlikely that such a high percentage of host particles is present in the observed M~31 W-shelf. Given the positions of the substructures vary with time in the simulations (see Section~\ref{sect:sim} for details), the presence of such large fractions of host particles in the two shelves (W-shelf certainly) implies that these substructures are simply at present too close to the M~31 disc in the comparison model snapshot. }

In \citet{Bh21}, we had also found that the NE-Shelf and GS PNLF shapes {connect their stellar population} to the {satellite} stellar population, {while they are quantitatively} different from that of the M~31 disc. The same conclusion was reached by \citet{Escala22} from the photometric metallicity of the NE-shelf and GS RGBs. The wedge patterns in the R$\rm_{proj}$ vs. LOSV phase space for the NE and W-shelves confirm the long-predicted hypothesis \citep{Ibata04} that the NE-shelf, W-shelf and GS were formed in the same merger event. However, given current observations these patterns only constrain the orbital characteristics of the infalling satellite and do not yet set definite values its mass and morphology (though deeper observations may help distinguish subtle differences).

\subsection{The Giant Stream and the infalling satellite}
\label{sect:gs}

%\begin{figure}
%	\centering
%	\includegraphics[width=\columnwidth,angle=0]{GS_host_sat.png}
%	\caption{Same as Figure~\ref{fig:proj} but for the simulated particles in the giant stream analogue only showing host particles [left] and satellite particles [right]. Overdensities (1ab -- red; 1bb -- cyan) in the projected phase space identified in the DESI survey are marked with dashed lines. }
%	\label{fig:proj_gs}
%\end{figure}

The broad LOSV distribution of the GS and its distribution in the R$\rm_{proj}$ vs. V$_{LOS}$ diagram as traced by PNe are consistent with the predictions from the major merger simulation of \citet{ham18}, whereas the minor merger scenario simulations predicted a narrower (20~km$\rm s^{-1}$) single component LOSV distribution as a function of R$\rm_{proj}$ \citep{Fardal12,Kirihara17}. %A fascinating prediction of the major merger simulation is that at low R$\rm_{proj}$, the GS contains not only  satellite debris but also a substantial amount of host particle debris (Figure~\ref{fig:proj_GS}). 

{A fascinating prediction of the major merger simulation is that at low R$\rm_{proj}$, the GS contains not only  satellite debris but also a substantial amount of host particle debris (see Figure~\ref{fig:proj_GS}).} In particular, note the central overdensities in the R$\rm_{proj}$ vs. V$_{LOS}$ diagram that are almost identical for both host (at low R$\rm_{proj}$) and satellite particles. The initial angular momentum of the host, satellite, and that of the GS {orbit} were all different from each other. So the only way for the satellite and host particles in the central overdensity to have the same angular momentum is that there are stars from the host now belonging to the GS. {The host particles were} likely brought along by the satellite during its multiple pericenter passages and deposited along the GS at low R$\rm_{proj}$. We note that the PNe sample at a higher R$\rm_{proj}$ region in the GS where their LOSV distribution is statistically distinct from that of the host particles (as found in Section~\ref{sect:los}) while the lower R$\rm_{proj}$ region is sampled by RGBs from \citet{Gilbert09}. 

There is also a clear wedge structure of host material (Figure~\ref{fig:proj_GS}~[left]). {The 1bb structure re-identified in \citet{Dey22} (originally called KCC by \citealt{Gilbert09}) out to large radii is consistent with being the edge of this wedge of host material. The 1ab structure (attributed to the main body of the GSS by \citealt{Gilbert09}) is consistent with a trail of satellite debris seen in Figure~\ref{fig:proj_GS}~[right].} {We reiterate that the bulk of satellite debris present in the central overdensity in this phase space has many corresponding RGB stars observed along this trail by \citet[][see their Figure 6]{Dey22} but were not marked as an overdensity by them.} The major merger model would thus explain the two kinematic components identified by \citet{Gilbert09} and \citet{Dey22} as dominated by host and satellite material respectively.

The origin of stars/PNe in the GS either from the host or satellite would be imprinted in their chemical abundances. Indeed a variation in the percentage mass of host material, decreasing with projected radial distance from M~31 along the GS, would naturally predict a variation of metallicity in the GS with R$\rm_{proj}$. A metallicity gradient along the GS has been measured from isochrone fitting of photometric RGB observations \citep[E.g.][]{Conn16} as well as {for} spectroscopy of RGB stars \citep{Escala21} though such a gradient was not observed by \citet{Dey22}. Comparison of the metallicity distributions of RGB stars with predictions from the major merger simulation by \citet{ham18} would be necessary to constrain the presence of host galaxy material in the GS, and is planned for a future work.

\subsection{Formation of an in-situ halo in M~31}
\label{sect:halo}

{Figure~\ref{fig:spat_sim} shows the number density map of simulated particles in the highest resolution major merger model of M~31 from \citet{ham18}. While host and/or satellite particles form the different inner halo substructures during the merger (as discussed in the previous sections), they also populate the extended halo in this simulation out to large radii. Indeed, in addition to the substructure velocities observed in the position-velocity plots in Figures~5--9%(\ref{fig:proj_G1}--\ref{fig:proj_G1})
, we observe particles (for both host and satellite) populating the entire range of velocities in each substructure region. These likely belong to the extended halo formed in the major merger simulation\footnote{Note that no halo particles were present at the start of the simulation; see Section~\ref{sect:sim} for details.}. \citet{ham18} have already shown that the mass density profile (see their Figure~14) and metallicity (see their Figure~16) of their simulated particles in the halo of M~31 is qualitatively consistent with that observed from RGB stars \citep[PAndAS,][]{ibata14}. Here we explore the relative fraction of host and satellite particles in the M~31 halo in this major-merger simulation.}  

\begin{figure}
	\centering
	\includegraphics[width=\columnwidth,angle=0]{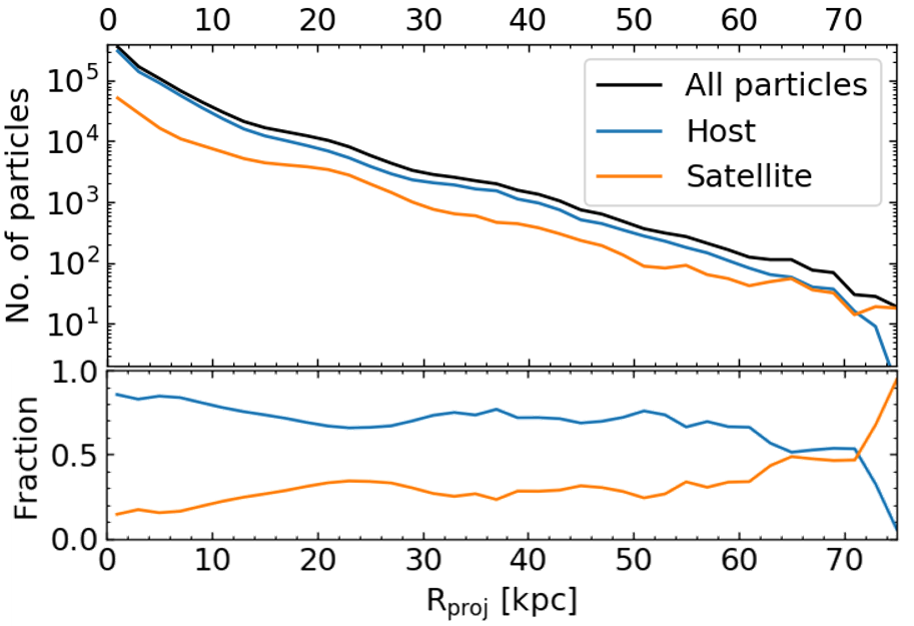}
	\caption{{[Top] The number of host (satellite) particles, in a 30$^{\circ}$ cone around the disc minor axis, as a function of projected radial distance (in bins of 2~kpc) from the center of the simulated M~31 model is shown in blue (orange). [Bottom] The same as above but for the fraction of host (satellite) particles.}}
	\label{fig:fraction}
\end{figure}

{Figure~\ref{fig:fraction} shows the number as well as fractions of host and satellite particles, in a 30$^{\circ}$ cone around the disc minor axis (to limit the contribution of the M~31 disc and exclude the GS), as a function of projected radial distance from the center of M~31. The fraction of satellite particles is low ($<20$\%) within R$\rm_{proj}\sim10$~kpc along the minor axis where the disc dominates but slowly increases to $\sim$35\% at R$\rm_{proj}\sim25$~kpc. At this radii, the NE- \& W-shelves are present in the simulation (see Figure~\ref{fig:spat_sim}) with a significant fraction of satellite debris (see Section 4.1). Beyond R$\rm_{proj}\sim30$~kpc, there are no bright substructures present and allows exploration of the inner halo of M~31 predicted from the simulation. The host particles constitute $\sim$70\% of the M~31 halo at R$\rm_{proj}\sim$30--60~kpc while the satellite particles constitute the other $\sim$30\%. Beyond R$\rm_{proj}\sim$60~kpc, the satellite particle fraction increases rapidly till R$\rm_{proj}\sim$75~kpc, beyond which almost no host particles are present and the sparse M~31 halo is dominated almost exclusively by satellite particles. The major merger simulation by \citet{ham18} thus predicts that the halo of M~31 is mainly constituted by in-situ host particles that dominate over the accreted satellite particles out to R$\rm_{proj}\sim$60~kpc. }

{We recall that in this simulation, the bulk of the satellite material is deposited in the central regions of M~31 (see Section~\ref{sect:sim}). This is consistent with the predictions from N-body simulations of \citet{Karademir19} where more massive satellites are found to deposit the bulk of their material at central regions of the host, whereas in minor mergers, the less massive satellites deposit their materials in the outskirts. }

{In the MW, the last major merger (also with mass ratio$\sim$1:4) has been reported to have happened $\sim$10~Gyr ago \citep[e.g.][]{Belokurov18, Helmi18}. This event is thought to have formed an additional inner halo component (called the splash by \citealt{Belokurov20}) from the pre-merger MW disc. \citet{Grand20} showed with the Auriga simulations of the MW that a major merger with a gas-rich satellite (almost $\sim$90\% gas mass given the early epoch) can form the puffed up MW thick disc and contribute an additional inner halo component from the particles in the proto-MW galaxy. }

{The major merger simulation (though at a much later epoch) for M~31 also predicts the formation of an in-situ halo from similarly splashed out host material that dominates the accreted halo from satellite material out to $\sim$60~kpc (Figure~\ref{fig:fraction}). This is qualitatively consistent with the identification of metal-rich RGB stars in the M~31 inner halo (in the disc line-of-sight as well as in pencil-beam fields in the inner halo substructure regions and along the minor axis of M~31) in the SPLASH survey \citep{Dorman13,Escala23} that are considered to be kicked-up stars from the M~31 disc stellar population. }

{A quantitative comparison of the observed number density \citep{Gilbert12,ibata14}, kinematic \citep{Dorman13,Gilbert12} and metallicity \citep{Escala20b,Wojno22,Escala23} measurements in M~31 with that of the extended halo from the major merger simulation by \citet{ham18} would be required to investigate if such a scenario is consistent with the observed halo properties in M~31. Such an investigation will be carried out in a forthcoming work.} 

{Finally, we note that given the ubiquity of host particles in the M~31 inner halo within R$\rm_{proj}\sim$60~kpc, it is unsurprising to find relatively high fractions of host material in the M~31 substructure fields studied in this work (see Table~\ref{table:prop}). The substructure spatial regions with relatively high fractions of satellite debris material, such as the GS and the two shelves, would thus be primarily composed of this satellite debris material but lying co-spatially on the host material-dominated inner halo, as has been discussed in Sections~\ref{sect:shelves}~\&~\ref{sect:gs}.}

\subsection{On the possibility that M~32 is the remnant core of the merged satellite}
\label{sect:m32}

{We note that given the NE-shelf, W-shelf and GS were formed in the same merger event, the orbit of the satellite is well constrained (see the orbital characteristics in \citealt{Fardal13} as well as in \citealt{ham18}) with the GS as the trailing trail of the satellite. \citet{dsouza18} presented arguments favouring the compact elliptical dwarf galaxy M~32 (projected 24$'$ away from the center of M~31; R$\rm_{proj}\sim5.4$~kpc) as the remnant of the infalling satellite by comparison with a candidate analogue from the Illustris cosmological simulation. However, in that analogue, the GS formed as a leading tail of the merger. Such an analogue is ruled out given the constraints on the M~31 satellite infall orbit with the GS being a trailing tail.} 

{In the major merger scenario from \citet{ham18}, which is favoured given the formation of the G1-Clump in the same event, the two merging galaxies are assumed to be bulgeless thin discs (see Section~\ref{sect:sim} for details). Given this assumption, no core would survive and most of the satellite material would be dispersed in the central regions of the host. However for the formation of a compact dwarf elliptical like M~32, the progenitor satellite would need to also have a dense bulge \citep{dsouza18}. The satellite would still be required to infall along the GS, produce the shelves, dynamically heat the M~31 thicker disc to its observed values \citep{Bh+19b} while also producing the G1-clump and its remant core would then need to be present at the observed position of M~32. Simulations similar to \citet{ham18} but with both disc galaxies having initial bulges would be required to constrain whether an M~32 like remnant is possible given the stringent orbit constrains. Furthermore, precise constraints on M~32's proper motion measurements (as is underway with HST; \citealt{Sohn20}) would allow us to determine if such satellite infall orbits are possible.}

\subsection{On the dark matter fraction of the satellite in merger simulations}
\label{sect:dm}
\citet{Sadoun14} and \citet{Milosevic22} support a similar minor-merger (mass ratio of barionic mass $\sim$1:20) scenario as \citet{Fardal13} but prefer a more massive satellite $\sim 4.4 \times 10^{10}$~M$_{\odot}$, including DM. \citet{Milosevic22} posit that the merger happened 2.4--2.9~Gyr ago which would produce a GS with two kinematic components (debris left in overlapping orbits of the satellite) that appears as a broad distribution of the GS in the R$\rm_{proj}$ vs. V$_{LOS}$ diagram, while also being consistent with the observed metallicity gradient of the GS \citep[E.g.][]{Conn16}. However, in their simulation the host galaxy has a DM mass fraction of 80\% while the satellite has a DM mass fraction of $\sim$95\% (note that in the simulations by \citealt{ham18}, both host and satellite had 80\% DM content). This means that though that their baryonic merger mass ratio may be $\sim$1:20 but the total merger mass ratio (including DM) is $\sim$1:5. Given the stellar mass -- halo mass relation favours higher DM fraction in dwarf galaxies \citep[E.g.][]{Brook14}, such a scenario is indeed possible. We however note that \citet{Milosevic22} had assumed a gas-free satellite infall and thus simulated a `dry' merger in M~31. The radial abundance gradient measured by PNe in the thin and thicker disc of M~31 \citep{Bhattacharya22} as well as their distribution in the log(O/Ar) vs. 12+ log(Ar/H) plane \citep{Arnaboldi22} necessitate a `wet' merger in M~31, ruling out a gas-free satellite.

Thus, we can not concretely claim that the infallen satellite in M~31 had a baryonic mass (stars as well as gas) $\sim$25\% that of M~31, but certainly its total mass (including DM) would be $\sim$25\% that of M~31. The formation of a G1-clump simulated counterpart, as well as keeping other constraints, in a merger simulation with different DM fractions for the host and satellite galaxies still needs to be explored. 

%______________________________________________________________

\section{Conclusion}
\label{sect:conc}
In this work, we present the first LOSV measurements of PNe co-spatial with five inner halo substructures of M~31. We compare their LOSV distribution and distribution in the R$\rm_{proj}$ vs. V$_{LOS}$ diagram with that from a major-merger simulation by \citet{ham18}. Our conclusions are as follows:
\begin{enumerate}
    %\item We identify the wedge structures in the R$\rm_{proj}$ vs. LOSV phase space for the W-shelf and NE-shelf, consistent with previous determinations of the latter from \citet{Escala22} using RGB stars. We determine the 3D structure of the NE- and W-shelves and concretely links them and the GS to the same merger event.
    \item The G1-clump is a dynamically cold substructure whose observed mean LOSV is remarkably consistent with being formed from the perturbed disc of M~31 ({independently supported from its PNLF; \citealt{Bh21}}) during the same major merger event that formed the GS, NE-shelf and W-shelf. It is the smoking gun for the major merger scenario (mass ratio $\sim$1:4 including DM; \citealt[][]{ham18}) in M~31 as such a structure can not be formed in a minor merger scenario. 
    \item {In the R$\rm_{proj}$ vs. V$_{LOS}$ diagrams for the two shelves and the GS, the PNe are consistent with the observations for their RGB counterparts and the predictions from the major merger simulations.  We identify the wedge structures in the R$\rm_{proj}$ vs. V$_{LOS}$ diagram for the W- and NE-shelves, while the GS PNe have an extended distribution in this plane.}
    \item {The major merger model by \citet{ham18} predicts different distributions of the host and satellite particles in the R$\rm_{proj}$ vs. V$_{LOS}$ diagrams for different substructures (see Section~\ref{sect:losvd}). In particular the GS is dominated by satellite material.} The two GS kinematic components, KCC and main GS, identified from RGBs \citep{Gilbert09, Dey22} are consistent with the host dominated and satellite dominated debris material respectively. However, the PNe co-spatial with the predicted CGS position, through their LOSVs, can neither validate nor {disprove} the existence of the CGS.
    \item {The major merger simulation also predicts that the M~31 inner (R$\rm_{proj}<60$~kpc) halo is primarily ($\sim$~70\%) constituted by in-situ host material that has been splashed out during the merger event. This is qualitatively consistent with the observed prevalence of kicked-up metal-rich disc RGB stars in the M~31 inner halo \citep{Escala23}}.
\end{enumerate}

{We can now paint a more comprehensive picture of the recent merger history of M~31 based on PN LOSV measurements, their PNLF \citep{Bh21}, the age velocity dispersion relation of the M~31 thin and thicker discs \citep{Bh+19b,dorman15} as well as their chemistry \citep{Arnaboldi22,Bhattacharya22}. M~31 has had a recent (2.5--4~Gyr ago\footnote{The PNe themselves would only limit the time of merger to 2.5--4.5~Gyr ago as the high-extinction PNe in the thin disc have a mean age of $\sim$2.5 Gyr while the low-extinction PNe in the thicker disc of M~31 have ages of 4.5~Gyr or more \citep{Bh+19b}. However, the presence of $\sim4$~Gyr old RGB stars in the GS \citep{Gilbert09} would set the time of the merger to between 2.5--4~Gyr ago.}) major merger with mass ratio $\sim$1:5 \citep{Bh+19b} with the gas-rich satellite falling along the Giant Stream that perturbed the pre-merger disc to form the M~31 thicker disc while also forming the NE- \& W-shelves from satellite debris, and the G1-clump from disc debris. The M~31 thin disc is re-formed in a burst of star formation \citep{bernard15,wil17} from pre-merger M~31 gas mixed with metal-poor gas brought in by the satellite \citep{Bhattacharya22,Arnaboldi22}. }

{The LOSV distributions of the M~31 inner halo substructures (G1-clump, NE and W-shelves, and GS) are interpreted consistently by the major merger simulation by \citet{ham18}. The G1-clump LOSV measurements in particular necessitate a recent major merger (including DM) in M~31 to form this substructure, ruling out minor merger scenarios (mass ratio $\sim$1:20--1:10 including DM). This wet merger simulation is also consistent with observations in its predictions of a disrupted thick disc in M~31 along with that of a thin disc from gas brought in by the satellite. }

{Next steps include a comparison of the major merger simulation predictions \citep{ham18} with RGB LOSV measurements from DESI \citep{Dey22} as well as with the chemistry of the PNe \citep{Bhattacharya22,Arnaboldi22} and RGB stars \citep{Escala23} in the M~31 disc and substructures respectively. Furthermore, deeper spectroscopic observations of substructure PNe would enable determination of their chemical abundances \citep[E.g.][]{Bhattacharya22}, and thereby the chemical abundances and enrichment history of the substructures they are associated with may be inferred \citep[as in][]{Arnaboldi22}. The subtle differences in the substructure phase space associated with different mass and morphologies of the infalling satellite \citep[E.g.][]{Kirihara17, ham18} would require a high resolution sampling of their phase space, possible when PNe and RGB LOSV measurements are utilized in conjunction.}

%LOSV measurements of the complete photometrically identified sample of PNe \citep{Bh21} in these substructures may improve the accuracy of their mean LOSV determination.   

%______________________________________________________________

\section*{Acknowledgements}
{We thank the anonymous referee for their comments.} SB is funded by the INSPIRE Faculty award (DST/INSPIRE/04/2020/002224), Department of Science and Technology (DST), Government of India. SB and MAR thank the European Southern Observatory, Garching, Germany for supporting SB's visit through the 2021 SSDF. {This research was supported by the Excellence Cluster ORIGINS which is funded by the Deutsche Forschungsgemeinschaft (DFG, German Research Foundation) under Germany´s Excellence Strategy – EXC-2094 – 390783311.} YY thanks the National Natural Foundation of China (NSFC No. 12041302 and No. 11973042). YY and FH thank for the support the International Research Program Tianguan, which is an agreement between the CNRS, NAOC and the Yunnan University. Based on observations obtained at the MMT Observatory, a joint facility of the Smithsonian Institution and the University of Arizona, and with MegaPrime/MegaCam, a joint project of CFHT and CEA/DAPNIA, at the Canada-France-Hawaii Telescope (CFHT). This research made use of Astropy-- a community-developed core Python package for Astronomy \citep{Rob13}, SciPy \citep{scipy}, NumPy \citep{numpy} and Matplotlib \citep{matplotlib}. This research also made use of NASA’s Astrophysics Data System (ADS\footnote{\url{https://ui.adsabs.harvard.edu}}).

%%%%%%%%%%%%%%%%%%%%%%%%%%%%%%%%%%%%%%%%%%%%%%%%%%
\section*{Data Availability}
Table~\ref{table:data} provides the required data on the  V$\rm_{LOS}$ of the PNe in each substructure. The PN spectra can be shared upon reasonable request to the authors.



\bibliographystyle{mnras}
\bibliography{ref_pne} % if your bibtex file is called example.bib

%\begin{thebibliography}{99}
%\end{thebibliography}

%%%%%%%%%%%%%%%%%%%%%%%%%%%%%%%%%%%%%%%%%%%%%%%%%%

%%%%%%%%%%%%%%%%% APPENDICES %%%%%%%%%%%%%%%%%%%%%
%\newpage
\appendix

% Don't change these lines
\bsp	% typesetting comment
\label{lastpage}
\end{document}